\documentclass{IEEEtran}
\usepackage{amsmath}
\usepackage{amsfonts}
\usepackage{amsthm}
\usepackage{amssymb}
\usepackage{cite}
\usepackage{url}
\usepackage{subfigure}
\usepackage{graphicx}
\usepackage{color}
\usepackage[ruled, lined,boxed,commentsnumbered]{algorithm2e}

\newcommand{\ie}{{\em i.e., }}

\newcommand{\bl}{\left(}
\newcommand{\br}{\right)}
\newcommand{\blc}{\left\{}
\newcommand{\brc}{\right\}}

\newcommand{\sgn}[1]{\text{sgn}\bl{#1}\br}

\newcommand{\beq} {\begin{equation}}
\newcommand{\eeq}[1] {\label{#1}\end{equation}}
\newcommand{\bfg} {\begin{figure}\begin{center}}
\newcommand{\efg}[1] {\end{center}\label{#1}\end{figure}}

\renewcommand{\vec}[1]{\mathbf{#1}}

\newcommand{\ba}{{\bf a}}
\newcommand{\bone}{{\bf 1}}
\newcommand{\bzero}{{\bf 0}}

\newcommand{\bs}{{\bf s}}
\newcommand{\bw}{{\bf w}}

\newcommand{\hw}{{\hat{\bf w}}}

\newcommand{\bt}{{\bf t}}

\newcommand{\bbr}{{\bf r}}

\newcommand{\bb}{{\bf b}}

\newcommand{\by}{{\bf y}}

\newcommand{\bg}{{\bf g}}

\newcommand{\be}{{\bf e}}

\newcommand{\bx}{{\bf x}}

\newcommand{\bd}{{\bf d}}

\newcommand{\bv}{{\bf v}}
\newcommand{\bu}{{\bf u}}
\newcommand{\bz}{{\bf z}}
\newcommand{\brr}{{\bf r}}

\newcommand{\la}{{\lambda}}

\newcommand{\bbbr}{{{\bf R}}}

\newcommand{\bbs}{{\bf S}}

\newcommand{\bbh}{{\bf H}}

\newcommand{\PP}{{\mathcal{P}}}

\newcommand{\Q}{{\mathcal{Q}}}

\newcommand{\B}{{\mathcal{B}}}
\newcommand{\A}{{\mathcal{A}}}
\newcommand{\G}{{\mathcal{G}}}
\newcommand{\C}{{\mathcal{C}}}
\newcommand{\OO}{{\mathcal{O}}}

\newtheorem{thm}{Theorem}

\title{Recursive $ \ell_{1,\infty} $ Group lasso}
\author{Yilun~Chen,~\IEEEmembership{Student Member,~IEEE,}
and~Alfred~O.~Hero,~III,~\IEEEmembership{Fellow,~IEEE}
\thanks{Y. Chen and A. O. Hero are with the Department of Electrical Engineering
and Computer Science, University of Michigan, Ann Arbor, MI 48109,
USA. Tel: 1-734-763-0564. Fax: 1-734-763-8041. Emails: \{yilun, hero\}@umich.edu.}

\thanks{This work was partially supported by AFOSR, grant number FA9550-06-1-0324.
}}

\begin{document}
\maketitle
\begin{abstract}
We introduce a recursive adaptive group lasso algorithm  for real-time penalized least squares prediction that produces a time sequence of optimal sparse predictor coefficient vectors. At each time index the proposed algorithm computes an exact update of the optimal $\ell_{1,\infty}$-penalized recursive least squares (RLS)  predictor. Each update minimizes a convex but non-differentiable function optimization problem. We develop an
on-line homotopy method to reduce the computational complexity. 
Numerical simulations demonstrate that the proposed algorithm outperforms the $\ell_1$
regularized RLS algorithm for a group sparse system identification problem and has lower implementation complexity than direct group lasso solvers.
\end{abstract}

\begin{IEEEkeywords}
RLS, group sparsity, mixed norm, homotopy, group lasso, system identification
\end{IEEEkeywords}

\section{Introduction}
 Recursive Least Squares (RLS)  is a widely used method for adaptive filtering and prediction in signal processing and related fields. Its applications include: acoustic echo cancelation; wireless channel equalization; interference cancelation and  data streaming predictors.
 In these applications a measurement stream is recursively fitted to a linear model, described by the coefficients of an FIR prediction filter, in such a way to minimize a weighted average of  squared residual prediction errors.
 Compared to other adaptive filtering algorithms such as Least Mean Square (LMS) filters, RLS is popular because of its fast convergence and low steady-state error.

In many applications it is natural to constrain the predictor coefficients to be sparse. In such cases the adaptive FIR prediction filter is a sparse system: only a few of the impulse response coefficients are non-zero. Sparse systems can be divided into general sparse systems and group sparse systems \cite{eldar2010block, chen2010regularized}. Unlike a general sparse system, whose impulse response can have arbitrary sparse structure, a group sparse system has impulse response composed of a few distinct clusters of non-zero coefficients. Examples of group sparse systems include specular multipath acoustic and wireless channels \cite{schreiber2002advanced, gu2009} and compressive spectrum sensing  of narrowband sources \cite{mishali2010theory}.

The exploitation of sparsity to improve prediction performance has attracted considerable interest. For general sparse systems, the $\ell_1$ norm has been recognized as an effective promotor of sparsity \cite{Tibshirani96, Candes06}. In particular, $\ell_1$ regularized LMS \cite{chen2009sparse, chen2010regularized} and RLS \cite{babadi2010sparls, angelosante2010online} algorithms have been proposed for for sparsification of adaptive filters. For group sparse systems, mixed norms such as the $\ell_{1,2}$ norm and the $\ell_{1,\infty}$ norm have been applied to promote sparsity in statistical regression \cite{yuan2006model, zhao2009composite, bach2008consistency}, commonly referred to as the group lasso, and sparse signal recovery in signal processing and communications \cite{eldar2010block, negahban2008joint}. However, most of the proposed estimation algorithms operate in the offline mode and are not designed for time varying systems and online prediction. This is the motivation of our work.

In this paper, we propose a RLS method penalized by the $\ell_{1,\infty}$ norm to promote group sparsity, called the recursive $\ell_{1,\infty}$ group lasso. Our recursive group lasso algorithm is suitable for online applications where data is acquired sequentially. The algorithm is based on the homotopy approach to solving the lasso problem and is an extension of \cite{garrigues2008homotopy, asif2010dynamic, malioutov2010sequential} to group sparse systems.

The paper is organized as follows. Section II formulates the problem. In Section III we develop the homotopy based algorithm to solve the recursive $\ell_{1,\infty}$ group lasso in an online recursive manner. Section IV provides numerical simulation results and Section V summarizes our principal conclusions. The proofs of theorems and some details of the proposed algorithm are provided in Appendix.

\emph{Notations}: In the following, matrices and vectors are denoted by boldface upper case letters and boldface lower case letters, respectively; $(\cdot)^T$ denotes the transpose operator,  and $\|\cdot\|_1$ and $\|\cdot\|_\infty$ denote the $\ell_1$ and $\ell_\infty$ norm of a vector, respectively;  for a set $\A$, $|\A|$ denotes its cardinality and $\phi$ denotes the empty set; $\bx_\A$ denotes the sub-vector of $\bx$ from the index set $\A$ and $\bbbr_{\A\B}$ denotes the sub-matrix of $\bbbr$ formed from the row index set $\A$ and column index set $\B$.

\section{Problem formulation}
\subsection{Recursive Least Squares}
Let $ \bw $ be a $ p $-dimensional coefficient vector. Let $ \by $ be an $ n $-dimensional vector comprised of observations $ \{y_j\}_{j=1}^n $. Let $\{{\mathbf x}_j\}_{j=1}^n $ be a sequence of $p$-dimensional predictor variables. In standard adaptive filtering terminology, $y_j$, ${\mathbf x}_j$ and $\mathbf w$ are the primary signal, the reference signal, and the adaptive filter weights. The RLS algorithm  solves the following quadratic minimization problem recursively over time $n=p,p+1,\ldots$:
\begin{equation}
	\label{eq:rls1}
\hw_n = \arg\min_{\bw}\sum_{j=1}^n \gamma^{n-j}(y_j-\bw^{T}\bx_j)^2,
\end{equation}
where $ \gamma \in (0,1]$ is the forgetting factor controlling the trade-off between transient and steady-state behaviors.

To serve as a template for the sparse RLS extensions described below we briefly review the RLS update algorithm.  Define $ \bbbr_n $ and $ \brr_n $ as
\begin{equation}
	\label{eq:Rn}
\bbbr_n = 	\sum_{j=1}^n \gamma^{n-j}\bx_j\bx_j^T
\end{equation}
and
\begin{equation}
	\label{eq:rn}
\bbr_n = \sum_{j=1}^n \gamma^{n-j} \bx_j y_j.
\end{equation}
The solution $ \hw_n $ to (\ref{eq:rls1}) can be then expressed as
\begin{equation}
	\label{eq:rls2}
\hw_n = \bbbr_n^{-1}\bbr_n.
\end{equation}
The matrix ${\mathbf R}_n$ and the vector ${\mathbf r}_n$ are updated as
\[
\bbbr_n = \gamma \bbbr_{n-1} + \bx_n\bx_n^{T},
\]
and
\[
\bbr_n = \gamma \bbr_{n-1} + \bx_n y_n^{T}.
\]
Applying the Sherman-Morrison-Woodbury formula \cite{hager1989updating},
\begin{equation}
	\label{eq:inverse_update}
\bbbr_n^{-1} = \gamma^{-1}\bbbr_{n-1}^{-1} - \gamma^{-1}\alpha_n \bg_n \bg_n^T,
\end{equation}
where
\begin{equation}
\bg_n = \bbbr_{n-1}^{-1}\bx_n
\end{equation}
and
\begin{equation}
\alpha_n = \frac{1}{\gamma + \bx_n^T\bg_n}.
\end{equation}
Substituting (\ref{eq:inverse_update}) into (\ref{eq:rls2}), we obtain the weight update \cite{Widrow85}
\begin{equation}
	\label{eq:rls3}
\hw_n = \hw_{n-1} + \alpha_n\bg_n e_n,
\end{equation}
where
\begin{equation}
    \label{eq:12}
e_n = y_n- \hw_{n-1}^{T}\bx_n.
\end{equation}
Equations (\ref{eq:inverse_update})-(\ref{eq:12}) define the RLS algorithm which has computational complexity of order $\mathcal O(p^2)$.

\subsection{Non-recursive $\ell_{1,\infty}$ group lasso}

\begin{figure}
\begin{center}
\includegraphics[scale=0.5]{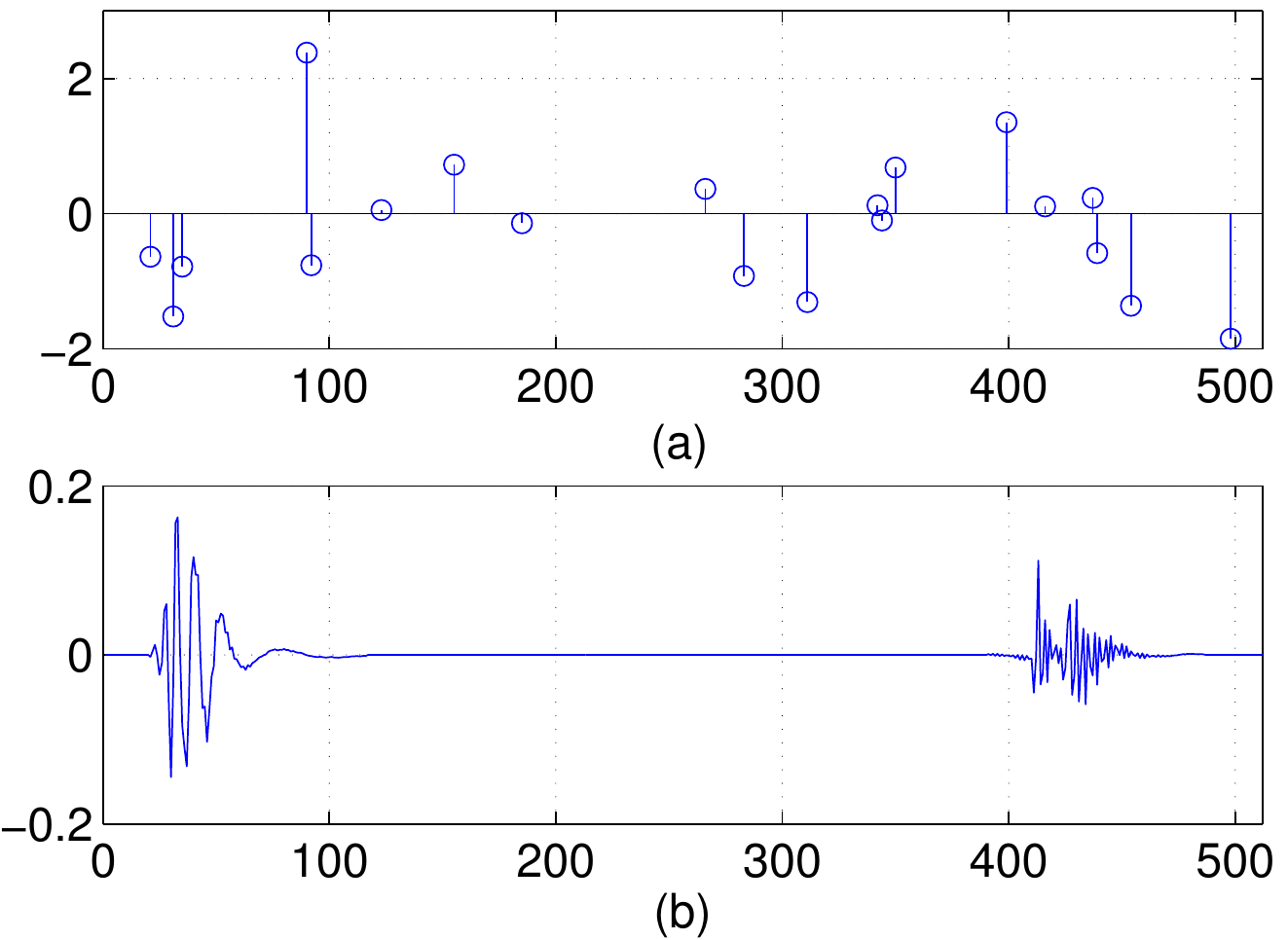}
\caption{Examples of (a) a general sparse system and  (b) a group-sparse system.}
\label{fig:0}
\end{center}
\end{figure}

The $ \ell_{1,\infty} $ group lasso is a regularized least squares approach which uses the $ \ell_{1,\infty} $ mixed norm to promote group-wise sparse pattern on the predictor coefficient vector. The $ \ell_{1,\infty} $ norm of a vector $ \bw $  is defined as
\[
\|\bw\|_{1,\infty} = \sum_{m=1}^{M} \|\bw_{\G_m}\|_\infty,
\]
where $ \{\G_m\}_{m=1}^M $ is a group partition of the index set $\G = \{1,\ldots,p\} $, \ie
\[
\bigcup_{m=1}^M \G_m = \G, \quad \G_m  \cap \G_{m'} = \phi \text{  if } m \ne m',
\]
and $ \bw_{\G_m} $ is a sub-vector of $ \bw $ indexed by $ \G_m $. The $ \ell_{1,\infty} $ norm is a mixed norm: it encourages correlation among coefficients inside each group
via the $\ell_\infty$ norm within each group and promotes sparsity across each group using the $\ell_1$ norm. The mixed norm $\|\bw\|_{1,\infty}$ is convex in $\bw$ and reduces to $\|\bw\|_1$ when each group contains only one coefficient, \ie
$|\G_1| = |\G_2| = \cdots = |\G_M| = 1.$

The $ \ell_{1,\infty} $ group lasso solves the following penalized least squares problem:
\begin{equation}
	\label{eq:rgl}
\hw_n = \arg\min_{\bw} \frac{1}{2} \sum_{j=1}^{n} \gamma^{n-j} (y_j-\bw^{T}\bx_j)^{2} + \lambda \|\bw\|_{1,\infty},
\end{equation}
where $ \lambda $ is a regularization parameter. Eq. (\ref{eq:rgl}) is a convex problem and can be solved by standard convex optimizers or path tracing algorithms \cite{zhao2009composite}. Direct solution of  (\ref{eq:rgl}) has computational complexity of $ \OO(p^3) $.

\subsection{Recursive $ \ell_{1,\infty} $ group lasso}
    \label{sec:subsecRGL}
In this subsection we obtain a recursive solution for (\ref{eq:rgl}) that gives an update $ \hw_n $ from $ \hw_{n-1} $. The approach taken is a group-wise generalization of recent works  \cite{garrigues2008homotopy, asif2010dynamic} that uses the homotopy approach to sequentially solve the lasso problem. Using the definitions (\ref{eq:Rn}) and (\ref{eq:rn}), the problem (\ref{eq:rgl}) is equivalent to
\begin{equation}
	\label{eq:rgl_2}
\begin{aligned}
\hw_n & = \arg\min_{\bw} \frac{1}{2} \bw^{T}\bbbr_{n}\bw -\bw^T\bbr_{n} + \lambda \|\bw\|_{1,\infty} \\
& = \arg\min_{\bw} \frac{1}{2} \bw^{T}\bl\gamma\bbbr_{n-1}+\bx_n^T\bx_n\br\bw \\
& \quad \quad -\bw^T(\gamma\bbr_{n-1}+\bx_n y_n) + \lambda \|\bw\|_{1,\infty}.
\end{aligned}
\end{equation}
Let $ f(\beta,\lambda) $ be the solution to the following parameterized problem
\begin{equation}
	\label{eq:rgl_3}
\begin{aligned}
f(\beta,\lambda) & = \arg\min_{\bw} \frac{1}{2} \bw^{T}\bl\gamma\bbbr_{n-1}+\beta\bx_n\bx_n^T\br\bw \\
& \quad \quad -\bw^T(\gamma\bbr_{n-1}+\beta\bx_n y_n) + \lambda \|\bw\|_{1,\infty}
\end{aligned}
\end{equation}
where $ \beta $ is a constant between 0 and 1. $ \hw_n $ and $ \hw_{n-1} $ of problem (\ref{eq:rgl_2}) can be expressed as
\[
\hw_{n-1} = f(0,\gamma\lambda),
\]
and
\[
\hw_n = f(1,\lambda).
\]
Our proposed method computes $ \hw_n $ from $ \hw_{n-1} $ in the following two steps:\\
\noindent
\textbf{Step 1}. Fix $ \beta = 0 $ and calculate $f(0,\lambda)$ from $f(0,\gamma\lambda)$. This is accomplished by computing the regularization path between $ \gamma\lambda $ and $ \lambda $ using homotopy methods introduced for the non-recursive $ \ell_{1,\infty} $ group lasso. The solution path is piecewise linear and the algorithm is described in \cite{zhao2009composite}.\\
\noindent
\textbf{Step 2}. Fix $ \lambda $ and calculate the solution path between $ f(0,\lambda) $ and $ f(1,\lambda) $. This is the key problem addressed in this paper.

To ease the notations we denote $ \bx_n $ and $ y_n $ by $ \bx $ and $ y $, respectively, and define the following variables:
\begin{equation}
    \label{eq:new_bbbr}
\bbbr(\beta) = \gamma \bbbr_{n-1} + \beta \bx\bx^{T}
\end{equation}
\begin{equation}
    \label{eq:new_brr}
\brr(\beta) = \gamma\brr_{n-1}+\beta\bx y.
\end{equation}
 Problem (\ref{eq:rgl_3}) is then
\begin{equation}
	\label{eq:homotopy_beta}
f(\beta,\la) = \arg \min_\bw \frac{1}{2}\bw^{T}\bbbr(\beta)\bw - \bw^T\brr(\beta) + \lambda \|\bw\|_{1,\infty}.
\end{equation}
In Section \ref{sec:homotopy_beta} we will show how to propagate $f(0,\lambda)$ to $ f(1,\lambda) $  using the homotopy approach applied to (\ref{eq:homotopy_beta}).

\section{Online homotopy update}
\label{sec:homotopy_beta}
\subsection{Set notation}
\label{sec:subsec_setnotation}
\begin{figure}
\centering
\includegraphics[width = 7cm]{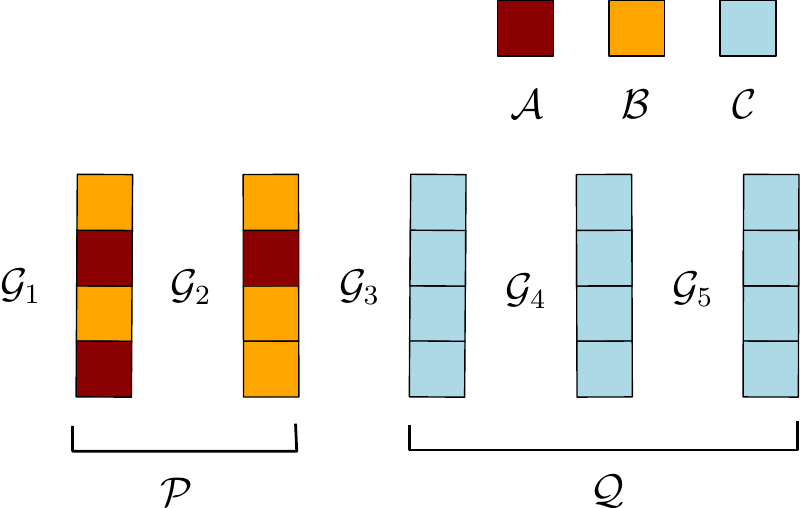}\\
  \caption{Illustration of the partitioning of a 20 element coefficient vector $\mathbf w$ into 5 groups of 4 indices. The sets $\mathcal P$ and $\mathcal Q$ contain the active groups and the inactive groups, respectively. Within each of the two active groups the maximal coefficients are denoted by the dark red color. }\label{fig:1}
\end{figure}

We begin by introducing a series of set definitions. Figure \ref{fig:1} provides an example.
We divide the entire group index set into $\PP$ and $Q$, respectively, where $\PP$ contains active groups and $\Q$ is its complement. For each active group $m \in \PP$, we partition the group into two parts: the maximal values, with indices $\mathcal A_m$,  and the rest of the values, with indices   $\mathcal B_m$:
\[
\A_m = \arg \max_{i \in \mathcal{G}_m} |w_i|, m \in \PP,
\]
and
\[
\B_m = \mathcal{G}_m - \A_m.
\]
The set $\A$ and $\B$ are defined as the union of the $ \A_m $ and $ \B_m $ sets, respectively:
\[
\A = \bigcup_{m \in \PP} \A_m, \quad  \B = \bigcup_{m \in \PP} \B_m.
\]
Finally, we define
\[
\C = \bigcup_{m \in \Q} \G_m.
\]
and
\[
  \C_m = \G_m \cap \C.
\]

\subsection{Optimality condition}
\label{sec:sub_opt_condition}
The objective function in (\ref{eq:homotopy_beta}) is convex but non-smooth as the $ \ell_{1,\infty} $ norm is non-differentiable. Therefore, problem (\ref{eq:homotopy_beta}) reaches its global minimum at $ \bw $ if and only if the sub-differential of the objective function contains the zero vector.  Let $ \partial \|\bw\|_{1,\infty} $ denote the sub-differential of the $ \ell_{1,\infty} $ norm at $ \bw $.  A vector $ \bz \in   \partial \|\bw\|_{1,\infty} $ only if $ \bz $ satisfies the following conditions \cite{negahban2008joint, zhao2009composite}:
\begin{align}
\label{eq:propty1}
&\|\bz_{\A_m}\|_1 = 1, m \in \PP,\\
\label{eq:propty2}
& \sgn{\bz_{\A_m}}  = \sgn{\bw_{\A_m}}, m \in \PP,\\
\label{eq:propty3}
&\bz_{\B} = \bf 0,\\
\label{eq:propty4}
&\|\bz_{\C_m}\|_1 \le 1, m \in \Q,
\end{align}
where $ \A,\B, \C, \PP $ and $ \Q $ are $\beta$-dependent sets defined on $ \bw $ as defined in Section \ref{sec:subsec_setnotation}.

For notational convenience we drop $\beta$ in $\mathbf R(\beta)$ and $\mathbf r(\beta)$  leaving the $\beta$-dependency  implicit. The optimality condition is then written as
\begin{equation}
\label{eq:kkt1}
\bbbr\bw - \brr + \la \bz = \bf 0, \quad \bz \in \partial \|\bw\|_{1,\infty}.
\end{equation}
As $ \bw_\C = \vec{0} $ and $ \bz_\B = \vec{0} $, (\ref{eq:kkt1})  implies the three conditions
\begin{align}
\label{eq:kkt21}
& \bbbr_{\A\A} \bw_\A + \bbbr_{\A\B} \bw_\B - \brr_\A + \la \bz_\A = \bf 0,\\
\label{eq:kkt22}
& \bbbr_{\B\A} \bw_\A + \bbbr_{\B\B} \bw_\B - \brr_\B = \bf 0,\\
\label{eq:kkt23}
& \bbbr_{\C\A} \bw_\A + \bbbr_{\C\B} \bw_\B - \brr_\C + \la \bz_\C = \bf 0.
\end{align}

The vector $ \bw_\A $ lies in a low dimensional subspace. Indeed, by definition of $\A_m$, if $|\mathcal A_m|>1$
\[
|w_i| = |w_{i'}|, ~~i, i' \in \A_m.
\]
Therefore, for any active group  $ m \in \PP $,
\begin{equation}
	\label{eq:wAm}
\bw_{\A_m} = \bs_{\A_m} \alpha_m
\end{equation}
where
\[
\alpha_m = \|\bw_{\G_m}\|_{\infty},
\]
and
\[
\bs_\A = \sgn{\bw_\A}.
\]
Using matrix notation, we represent (\ref{eq:wAm}) as
\begin{equation}
	\label{eq:wA}
\bw_\A = \bbs \ba.
\end{equation}
where
\begin{equation}
    \label{eq:def_bbs}
{\bf{S}} = \left( {\begin{array}{*{20}{c}}
   {{{\bf{s}}_{{A_1}}}} & {} & {}  \\
   {} &  \ddots  & {}  \\
   {} & {} & {{{\bf{s}}_{{A_{|P|}}}}}  \\
\end{array}} \right)
\end{equation}
is a $ |\A| \times |\PP|$ sign matrix and the vector $ \ba $ is comprised of $\alpha_m, m\in \PP $.

The solution to (\ref{eq:homotopy_beta}) can be determined in closed form if the sign matrix $\bbs$ and sets $(\A,\B,\C,\PP,\Q)$ are available. Indeed, from (\ref{eq:propty1}) and (\ref{eq:propty2})
\begin{equation}
    \label{eq:StZa}
    \bbs^T \bz_\A = \bone,
\end{equation}
where $\bone$ is a $|\PP| \times 1$ vector comprised of 1's. With (\ref{eq:wA}) and (\ref{eq:StZa}), (\ref{eq:kkt21}) and (\ref{eq:kkt22}) are equivalent to
\begin{equation}
    \label{eq:kkt3}
  \begin{aligned}
   & \bbs^T\bbbr_{\A\A}\bbs \ba + \bbs^T\bbbr_{\A\B}\bw_\B - \bbs^T\brr_\A + \la\bone = 0,\\
   & \bbbr_{\B\A}\bbs\ba + \bbbr_{\B\B}\bw_\B-\brr_\B = \bzero.
  \end{aligned}
\end{equation}
Therefore, by defining the (a.s. invertible) matrix
\begin{equation}
    \label{eq:def_bbh}
{\bf{H}} = {
\begin{pmatrix}
   {{{\bf{S}}^T}{{\bf{R}}_{\A\A}}{\bf{S}}} & {{{\bf{S}}^T}{{\bf{R}}_{\A\B}}}  \\
   {{{\bf{R}}_{\B\A}}{\bf{S}}} & {{{\bf{R}}_{\B\B}}}  \\
\end{pmatrix}},
\end{equation}
and
\begin{equation}
    \label{eq:def_v}
{\bf{b}} = \begin{pmatrix}
   {{{\bf{S}}^T}{{\bf{r}}_\A}}  \\
   {{{\bf{r}}_\B}}  \\
\end{pmatrix},
{\bf{v}} = \left( {\begin{array}{*{20}{c}}
   {\bf{a}}  \\
   {{{\bf{w}}_B}}  \\
\end{array}} \right),
\end{equation}
 (\ref{eq:kkt3}) is equivalent to
$\bbh \bv = \bb - \la \be$, where $\be = (\mathbf 1^T, \mathbf 0^T)^T$, so that
\begin{equation}
    \label{eq:direct_solution}
  \bv = \bbh^{-1}(\bb-\la\be).
\end{equation}
 As $\bw_\A = \bbs \ba$, the solution vector $\bw$ can be directly obtained from $\bv$ via (\ref{eq:def_v}). For the sub-gradient vector,  it can be shown that
\begin{equation}
    \label{eq:def_bzA}
  \la\bz_\A = \brr_\A - \bl\bbbr_{\A\A}\bbs~~\bbbr_{\A\B}\br \bv,
\end{equation}
\begin{equation}
  \bz_\B = \bzero
\end{equation}
and
\begin{equation}
    \label{eq:def_bzC}
  \la\bz_\C = \brr_\C - \bl\bbbr_{\C\A}\bbs~~\bbbr_{\C\B}\br \bv.
\end{equation}

\subsection{Online update}
Now we consider (\ref{eq:homotopy_beta}) using the results in \ref{sec:sub_opt_condition}. Let $\beta_0$ and $\beta_1$ be two constants such that $\beta_1 > \beta_0$. For a given value of $\beta\in [\beta_0,\beta_1]$ define the class of sets ${\mathcal S}= (\A,\B,\C,\PP,\Q)$ and make $\beta$ explicit by writing ${\mathcal S}(\beta)$. Recall that ${\mathcal S}(\beta)$ is specified by the solution $f(\beta,\lambda)$ defined in (19). Assume that ${\mathcal S}(\beta)$ does not change for $\beta \in [\beta_0,\beta_1]$. The following theorem propagates $f(\beta_0,\la)$ to $f(\beta_1,\la)$ via a simple algebraic relation.
\begin{thm}
    \label{thm:1}
    Let $\beta_0$ and $\beta_1$ be two constants such that $\beta_1 >\beta_0$ and for any $\beta \in [\beta_0,\beta_1]$ the solutions to (\ref{eq:homotopy_beta}) share the same sets $\mathcal S = (\A,\B,\C,\PP,\Q)$.  Let $\bv'$ and $\bv$ be vectors defined as $f(\beta_1,\la)$ and $f(\beta_0,\la)$, respectively. Then
    \begin{equation}
        \label{eq:update_v}
      \bv' = \bv + \frac{\beta_1-\beta_0}{1+\sigma_H^2\beta_1}(y-\hat y) \bg,
    \end{equation}
and the corresponding sub-gradient vector has the explicit update
    \begin{equation}
	\label{eq:update_zA}
\begin{aligned}
 \lambda {\bf{z}}_\A' = \lambda {{\bf{z}}_\A} + \frac{\beta_1-\beta_0}{1+\sigma_H^2\beta_1}\left( {y - \hat y} \right)\left\{ {{{\bf{x}}_\A} -
(\bbbr_{\A\A}\bbs~~\bbbr_{\A\B})
\bg} \right\}
\end{aligned}
\end{equation}
and
\begin{equation}
	\label{eq:update_zC}
\begin{aligned}
\lambda {\bf{z}}_\C'
 = \lambda {{\bf{z}}_\C} + \frac{\beta_1-\beta_0}{1+\sigma_H^2\beta_1}\left( {y - \hat y} \right)\left\{ {{{\bf{x}}_\C} -
( \bbbr_{\C\A}\bbs~~\bbbr_{\C\B})
\bg} \right\},
\end{aligned}
\end{equation}
where $\bbbr = \bbbr(0)$ as defined in (\ref{eq:new_bbbr}), $(\bx,y)$ is the new sample as defined in (\ref{eq:new_bbbr}) and (\ref{eq:new_brr}),  the sign matrix $\bbs$ is obtained from the solution at $\beta = \beta_0$,  $\bbh_0$ is calculated from (\ref{eq:def_bbh}) using $\bbs$ and $\bbbr(0)$,  and $\bd$, $\bu$, $\hat y$ and $\sigma_H^2$ are defined by
\begin{equation}
    \label{eq:def_bd}
{\bf{d}} = \left( {\begin{array}{*{20}{c}}
   {{{\bf{S}}^T}{{\bf{x}}_\A}}  \\
   {{{\bf{x}}_\B}}  \\
\end{array}} \right),
\end{equation}
\begin{equation}
    \label{eq:def_bg}
  \bg = \bbh_0^{-1}\bd,
\end{equation}
\begin{equation}
    \label{eq:def_hat_y}
  \hat y = \bd^T \bv,
\end{equation}
\begin{equation}
    \label{eq:def_sigmaH2}
  \sigma_H^2 = \bd^T\bg.
\end{equation}
\end{thm}
The proof of Theorem \ref{thm:1} is provided in Appendix A. Theorem \ref{thm:1} provides the closed form  update for the solution path $f(\beta_0,\la) \rightarrow f(\beta_1,\la)$,  under the assumption that the associated sets $\mathcal S(\beta)$ remain unaltered over the path.

Next, we partition the range $\beta \in [0,1]$ into contiguous segments over which $\mathbf S(\beta)$ is piecewise constant. Within each segment we can use Theorem 1 to propagate the solution from left endpoint to right endpoint. Below we specify an algorithm for finding the endpoints of each of these segments.

Fix an endpoint $\beta_0$ of one of these segments. We seek a \emph{critical point} $\beta_1$ that is defined as the maximum $\beta_1$ ensuring $\mathcal S(\beta)$ remains unchanged within $[\beta_0,\beta_1]$. By increasing $\beta_1$ from $\beta_0$, the sets $\mathcal S(\beta)$ will not change until at least one of the following conditions are met:\\
\noindent\textbf{Condition 1}. There exists $i \in \A$ such that $ z_i' = 0$;\\
\noindent\textbf{Condition 2}. There exists $i \in \B_m$ such that $|w_i'| = \alpha_m'$;\\
\noindent\textbf{Condition 3}. There exists $m \in \PP$ such that $\alpha_m' = 0$;\\
\noindent\textbf{Condition 4}. There exists $m \in \Q$ such that $\|\bz_{\C_m}'\|_1 = 1$.\\
Condition 1 is from (\ref{eq:propty2}) and (\ref{eq:propty3}), Condition 2 and 3 are based on definitions of $\A$ and $\PP$, respectively, and Condition 4 comes from (\ref{eq:propty1}) and (\ref{eq:propty4}). Following \cite{efron2004least, zhao2009composite}, the four conditions can be assumed to be mutually exclusive. The actions with respect to Conditions 1-4 are given by\\
\noindent\textbf{Action 1}. Move the entry $i$ from $\A$ to $\B$:
\[
\A \leftarrow \A - \{i\}, \B \leftarrow \B \cup \{i\};
\]
\noindent\textbf{Action 2}. Move the entry $i$ from $\B$ to $\A$:
\[
\A \leftarrow \A \cup \{i\}, \B \leftarrow \B - \{i\};
\]
\noindent\textbf{Action 3}. Remove group $m$ from the active group list
\[
  \PP \leftarrow \PP - \{m\}, \Q \leftarrow \Q \cup \{m\},
\]
and update the related sets
\[
\A \leftarrow \A - \A_m, \C \leftarrow \C \cup \A_m;
\]
\noindent\textbf{Action 4}. Select group $m$
\[
  \PP \leftarrow \PP \cup \{m\}, \Q \leftarrow \Q - \{m\},
\]
and update the related sets
\[
\A \leftarrow \A \cup \C_m, \C \leftarrow \C - \C_m.
\]


By Theorem \ref{thm:1}, the solution update from $\beta_0$ to $\beta_1$ is in closed form. The critical point of $\beta_1$ can be determined in a straightforward manner (details are provided in Appendix B). Let $\beta_{1}^{(k)}, k=1,...,4$ be the minimum value that is greater than $\beta_0$ and meets Condition 1-4, respectively. The critical point $\beta_1$ is then
\[
  \beta_1 = \min_{k=1,...,4}\beta_1^{(k)}.
\]

\subsection{Homotopy algorithm implementation}
\label{sec:sub_algo_complexity}
We now have all the ingredients for the homotopy update algorithm and the pseudo code is given  in Algorithm \ref{alg:1}.
\begin{algorithm}
\label{alg:1}
\SetKwInOut{Input}{Input}\SetKwInOut{Output}{output}
\Input{$f(0,\la), \bbbr(0), \bx, \by$}
\Output{$f(1,\la)$}
\BlankLine

Initialize $\beta_0 = 0$, $\beta_1 =  0$, $\bbbr = \bbbr(0)$\;
Calculate $(\A,\B,\C,\PP,\Q)$ and $(\bv, \la\bz_\A, \la\bz_\C)$ from $f(0,\la)$\;
\While{$\beta_0 < 1$}
{
    Calculate the environmental variables $(\bbs, \bbh_0, \bd, \bg, \hat y, \sigma_H^2)$ from $f(\beta_0,\la)$ and $\bbbr$\;
    Calculate $\{\beta_1^{(k)}\}_{k=1}^4$ that meets Condition 1-4, respectively\;
    Calculate the critical point $\beta_1$ that meets Condition $k_*$:
    $k_* = \arg \min_{k} \beta_1^{(k)}$ and $\beta_1 = \beta_1^{(k_*)}$\;
    \eIf{$\beta_1 \le 1$}
    {
        Update $(\bv, \la\bz_\A, \la\bz_\C)$ using (\ref{eq:update_v}), (\ref{eq:update_zA}) and (\ref{eq:update_zC})\;
        Update $(\A,\B,\C,\PP,\Q)$ by Action $k_*$\;
        $\beta_0 = \beta_1$\;
    }
    {
        \textbf{break}\;
    }
}
$\beta_1 = 1$\;
Update $(\bv, \la\bz_\A, \la\bz_\C)$ using (\ref{eq:update_v})\;
Calculate $f(1,\la)$ from $\bv$.
  \caption{Homotopy update from $f(0,\la)$ to $f(1,\la)$.}
\end{algorithm}

Next we analyze the computational cost of Algorithm \ref{alg:1}. The complexity to compute each critical point is summarized in Table \ref{tab:1}, where $N$ is the dimension of $\bbh_0$. As $N = |\PP|+|\B| \le |\A|+|\B|$, $N$ is upper bounded by the number of non-zeros in the solution vector.  The vector $\bg$ can be computed in $\OO(N^2)$ time using the matrix-inverse lemma \cite{hager1989updating} and the fact that, for each action, $\bbh_0$ is at most perturbed by a rank-two matrix. This implies that the computation complexity per critical point is $\OO(p \max\{N,\log p\})$ and the total complexity of the online update  is $\OO(k_2\cdot  p \max\{N,\log p\})$, where $k_2$ is the number of critical points of $\beta$ in the solution path $f(0,\la) \rightarrow f(1,\la)$. This is the computational cost required for Step 2 in Section \ref{sec:subsecRGL}.

A similar analysis can be performed for the complexity of Step 1, which requires $\OO(k_1 \cdot p \max\{N,\log p\})$ where $k_1$ is the number of critical points in the solution path $f(0,\gamma\la) \rightarrow f(0,\la)$. Therefore, the overall computation complexity of the recursive $\ell_{1,\infty}$ group lasso is $\OO(k \cdot p \max\{N,\log p\})$, where $k = k_1 + k_2$, \ie the total number of critical points in the solution path $f(0,\gamma\la) \rightarrow f(0,\la) \rightarrow f(1,\la)$.

 An instructive benchmark is to directly solve the $n$-samples problem  (\ref{eq:rgl_3}) from the solution path $f(1,\infty)$ (\ie a zero vector) $\rightarrow f(1,\lambda)$ \cite{zhao2009composite}, without using the previous solution $\hw_{n-1}$. This algorithm, called iCap in \cite{zhao2009composite}, requires $\OO(k' \cdot p \max\{N,\log p\})$, where $k'$ is the number of critical points in $f(1,\infty) \rightarrow f(1,\lambda)$. Empirical comparisons between $k$ and $k'$, provided in the following section, indicate that iCap requires significantly more computation than our proposed Algorithm \ref{alg:1}.

\begin{table}
\centering
\begin{tabular}{|c|c|}
\hline
$\bg = \bbh_0^{-1}\bd$ & $\OO(N^2)$\\
\hline
$\bx_\A-(\bbbr_{\A\A}\bbs~~\bbbr_{\A\B})\bg$ & $\OO(|\A| N)$\\
\hline
$\bx_\C-(\bbbr_{\C\A}\bbs~~\bbbr_{\C\B})\bg$ & $\OO(|\C| N)$\\
\hline
$\beta_1^{(1)}$ & $\OO(|\A|)$\\
\hline
$\beta_1^{(2)}$ & $\OO(|\B|)$\\
\hline
$\beta_1^{(3)}$ & $\OO(|\PP|)$\\
\hline
$\beta_1^{(4)}$ & $\OO(|\C|\log|\C|)$\\
\hline
\end{tabular}
\caption{Computation costs of online homotopy update for each critical point.}
\label{tab:1}
\end{table}

\section{Numerical simulations}

\begin{figure}
\centering
\includegraphics[width = 7cm]{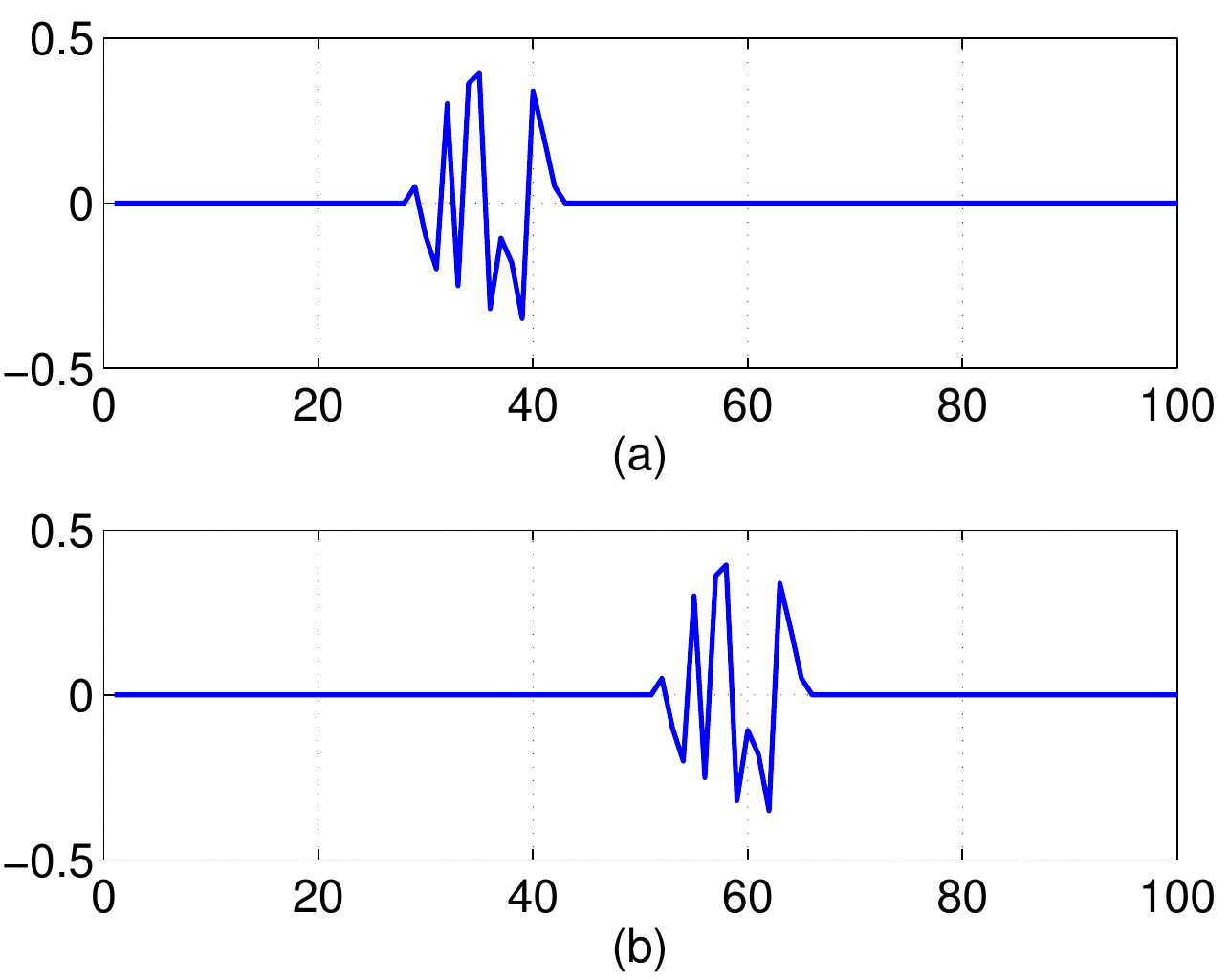}\\
  \caption{Responses of the time varying system. (a): Initial response. (b): Response after the 200th iteration. The groups for Algorithm 1 were chosen as 20 equal size contiguous groups of coefficients partitioning the range $1, \ldots, 100$.}\label{fig:2}
\end{figure}

In this section we demonstrate our proposed recursive $\ell_{1,\infty}$ group lasso algorithm by numerical simulation. We simulated the model $y_j = {\mathbf w}_*^T {\mathbf x}_j+ v_j$, $j=1, \ldots, 400$, where $v_j$ is a zero mean Gaussian noise and ${\mathbf w}_*$ is a sparse $p=100$ element vector containing only 14 non-zero coefficients clustered between indices 29 and 42.  See Fig. \ref{fig:2} (a). After 200 time units, the locations of the non-zero coefficients of ${\mathbf w}_*$ is shifted to the right, as indicated in Fig. \ref{fig:2} (b).

\begin{figure}
\centering
\includegraphics[width = 7cm]{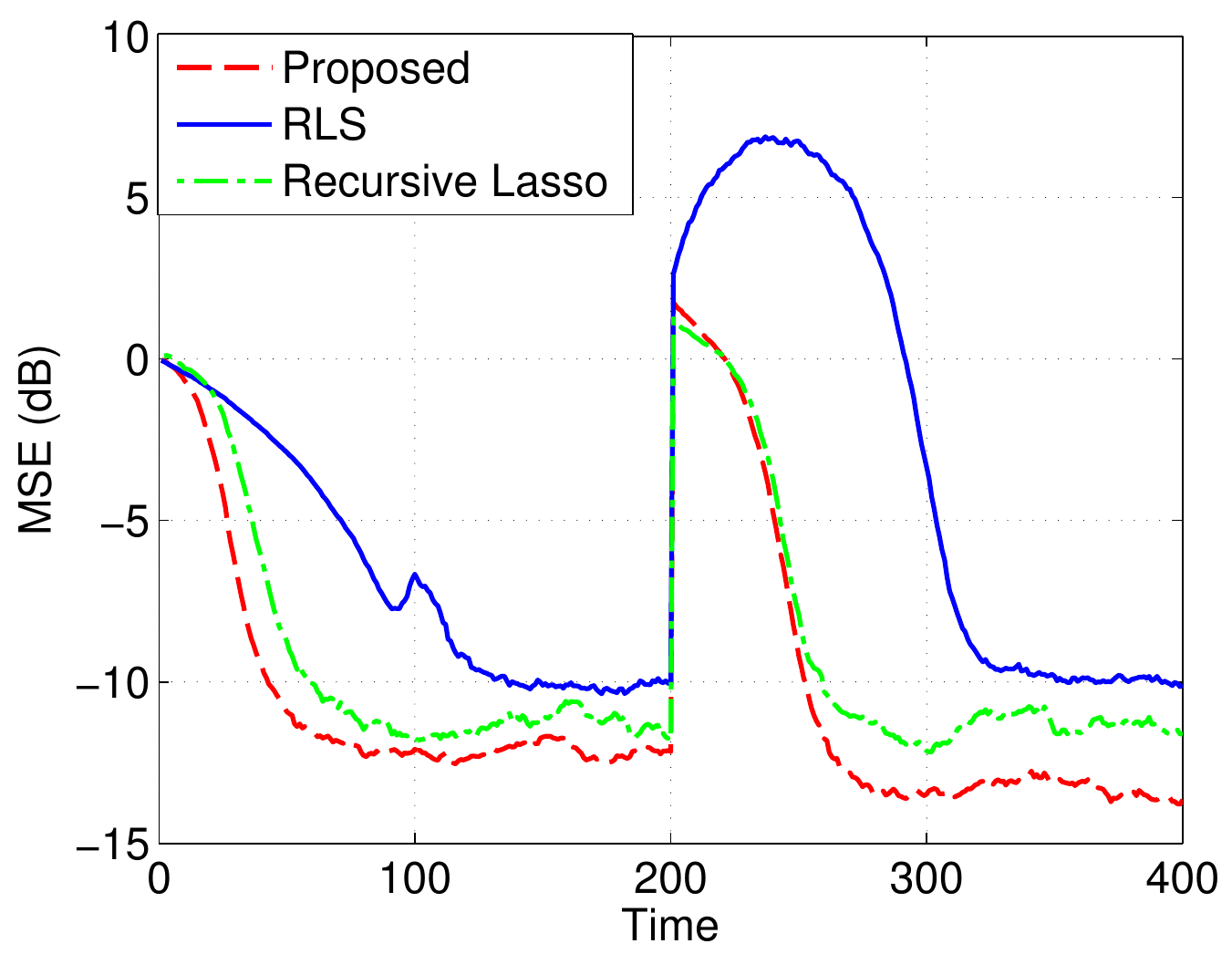}\\
  \caption{Averaged MSE of the proposed algorithm, RLS and recursive lasso.}\label{fig:3}
\end{figure}

The input vectors were generated as independent identically distributed Gaussian random vectors with zero mean and identity covariance matrix, and the variance of observation noise $v_j$ is 0.01. We created the groups in the recursive $\ell_{1,\infty}$ group lasso as follows. We divide the 100 RLS filter coefficients $\mathbf w$ into 20 groups with group boundaries $1, 5, 10, \ldots$, where each group contains 5 coefficients. The forgetting factor $\gamma$ and the regularization parameter $\lambda$ were set to 0.9 and 0.1, respectively. We repeated the simulation 100 times and the averaged mean squared errors of the RLS, sparse RLS and proposed RLS shown in Fig. \ref{fig:3}. We implemented the standard RLS and sparse RLS using the $\ell_1$ regularization, where the forgetting factors are also set to 0.9. We implemented sparse RLS \cite{garrigues2008homotopy} by choosing the regularization parameter $\lambda$ which achieves the lowest steady-state error, resulting in $\lambda=0.05$.

It can be seen from Fig. \ref{fig:3} that our proposed sparse RLS method outperforms standard RLS and sparse RLS in both convergence rate and steady-state MSE. This demonstrates the power of our group sparsity penalty. At the change point of 200 iterations, both the proposed method and sparse RLS of \cite{garrigues2008homotopy} show superior tracking performances as compared to the standard RLS. We also observe that the proposed method achieves even smaller MSE after the change point occurs. This is due to the fact that the active cluster spans across group boundaries in the initial system  (Fig. \ref{fig:2} (a)), while the active clusters in the shifted system overlap with fewer groups.

\begin{figure}
\centering
\includegraphics[width = 7cm]{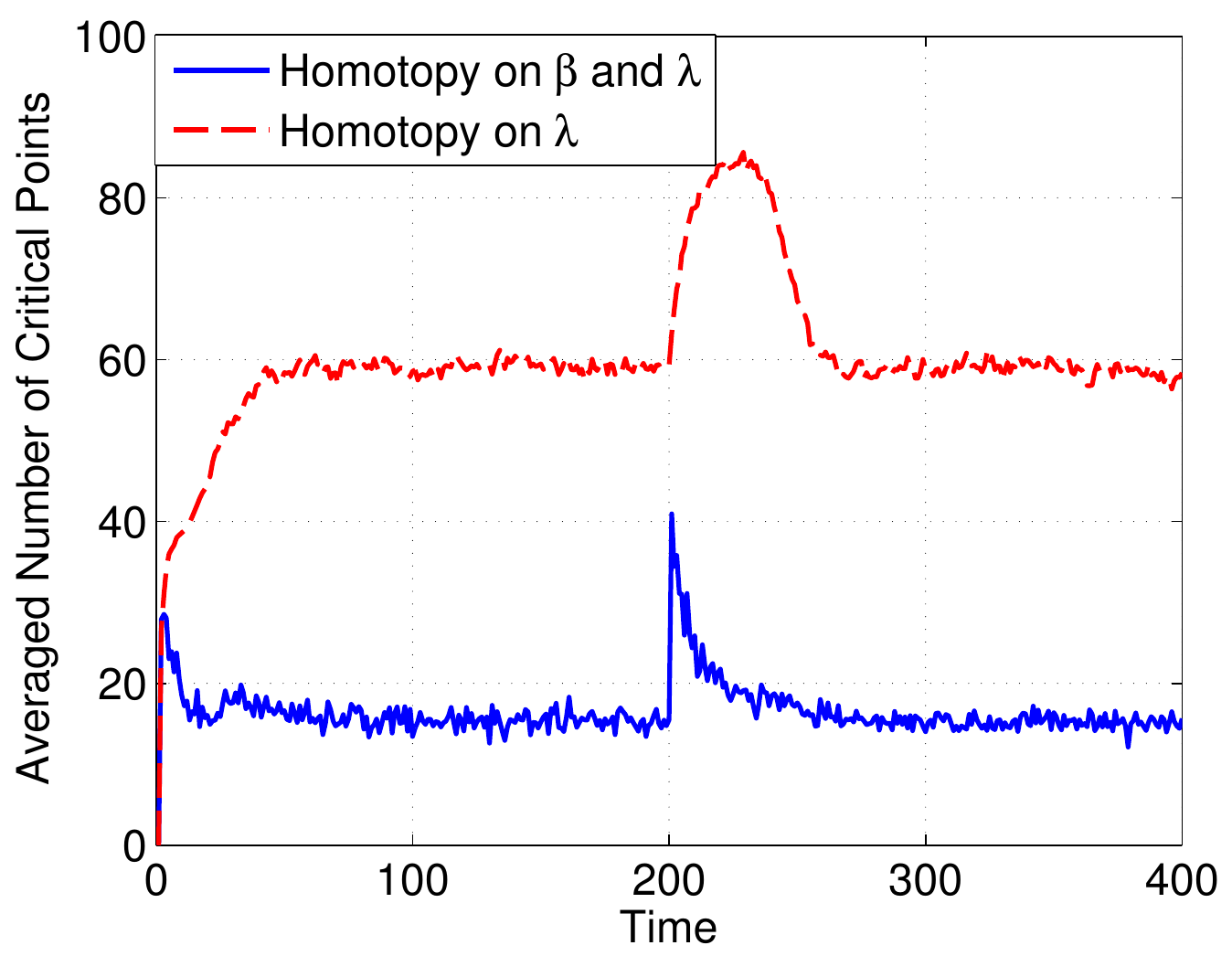}\\
  \caption{Averaged number of critical points for the proposed recursive method of implementing $\ell_{1,\infty}$ lasso and the  iCap \cite{zhao2009composite} non-recursive method of implementation.}\label{fig:4}
\end{figure}

Fig. \ref{fig:4} shows the average number of critical points (accounting for  both trajectories in $\beta$ and $\la$) of the proposed algorithm, \ie the number $k$ as defined in Section \ref{sec:sub_algo_complexity}. As a comparison, we implement the iCap method of \cite{zhao2009composite}, a homotopy based algorithm that traces the solution path only over $\lambda$.  The average number of critical points for iCap is plotted in Fig. \ref{fig:4}, which is the number $k'$ in Section \ref{sec:sub_algo_complexity}. Both the proposed algorithm and iCap yield the same solution but have different computational complexities proportional to  $k$ and $k'$, respectively. It can be seen that the proposed algorithm saves as much as 75\% of the computation costs  for equivalent performance.

\section{Conclusion}
In this paper we proposed a $\ell_{1,\infty}$ regularized RLS algorithm for online sparse linear prediction. We developed a homotopy based method to sequentially update the solution vector as new measurements are acquired. Our proposed algorithm uses the previous estimate as a ``warm-start", from which we compute the homotopy update to the current solution. The proposed algorithm can process streaming measurements with time varying predictors and is computationally efficient compared to non-recursive group lasso solvers. Numerical simulations demonstrated that the proposed method outperformed the standard and $\ell_1$ regularized RLS for identifying an unknown group sparse system, in terms of both tracking and steady-state mean squared error.

The work presented here assumed non-overlapping group partitions. In the future, we will investigate overlapping groups and other flexible partitions \cite{jenatton2009structured}.
\section{Appendix}
\subsection{Proof of Theorem \ref{thm:1}}
We begin by deriving (\ref{eq:update_v}). According to (\ref{eq:direct_solution}),
\begin{equation}
    \label{eq:app70}
  \bv' = \bbh'^{-1}(\bb'-\la\be').
\end{equation}
As $\bbs$ and $(\A,\B,\C,\PP,\Q)$ remain constant within $[\beta_0,\beta_1]$,
\begin{equation}
    \label{eq:app71}
\be' = \be,
\end{equation}
\begin{equation}
    \label{eq:app72}
  \bb' = \bb +
  \delta \bd y,
\end{equation}
and
\[
\bbh' = \bbh + \delta \bd\bd^T,
\]
where
\[
  \delta = \beta_1 - \beta_0,
\]
$\bbh$ and $\bb$ are calculated using $\bbs$ within $[\beta_0,\beta_1]$ and $\bbbr(\beta_0)$ and $\brr(\beta_0)$, respectively. We emphasize that $\bbh$ is based on $\bbbr(\beta)$ and is different from $\bbh_0$ defined in Theorem \ref{thm:1}. According to the Sherman-Morrison-Woodbury formula,
\begin{equation}
    \label{eq:app75}
  \bbh'^{-1} = \bbh^{-1} - \frac{\delta}{1+\sigma^2 \delta} (\bbh^{-1}\bd)(\bbh^{-1}\bd)^T,
\end{equation}
where $\sigma^2 = \bd^T \bbh^{-1}\bd$. Substituting (\ref{eq:app71}), (\ref{eq:app72}) and (\ref{eq:app75}) into (\ref{eq:app70}), after simplification we obtain
\begin{equation}
    \label{eq:app76}
  \begin{aligned}
  \bv' &= \bl\bbh^{-1} - \frac{\delta}{1+\sigma^2 \delta} (\bbh^{-1}\bd)(\bbh^{-1}\bd)^T\br \bl \bb + \delta\bd y - \la \be\br\\
  & = \bbh^{-1}(\bb-\la\be)  + \bbh^{-1}\delta\bd y \\
  & ~~~~~-\frac{\delta}{1+\sigma^2\delta}\bbh^{-1}\bd\bd^T\bbh^{-1}(\bb-\la\be) - \frac{\sigma^2\delta^2}{1+\sigma^2\delta}\bbh^{-1}\bd y \\
  & = \bv + \frac{\delta}{1+\sigma^2\delta} (y-\bd^T\bv) \bbh^{-1}\bd \\
  & = \bv + \frac{\delta}{1+\sigma^2\delta} (y-\hat y) \bbh^{-1}\bd,
  \end{aligned}
\end{equation}
where $\hat y = \bd^T\bv$ as defined in (\ref{eq:def_hat_y}).

Note that $\bbh$ is defined in terms of $\bbbr(\beta_0)$ rather than $\bbbr(0)$ and
\[
  \bbh = \bbh_0 + \beta_0 \bd\bd^T,
\]
so that
\begin{equation}
    \label{eq:app78}
  \bbh^{-1} = \bbh_0^{-1} - \frac{\beta_0}{1+\sigma_H^2 \beta_0} \bg\bg^T,
\end{equation}
where $\bg$ and $\sigma_H^2$ are defined by (\ref{eq:def_bg}) and (\ref{eq:def_sigmaH2}), respectively. As $\sigma_H^2 = \bd^T\bg$,
\begin{equation}
    \label{eq:app79}
  \bbh^{-1}\bd = \bbh_0^{-1}\bd - \frac{\sigma_H^2\beta_0}{1+\sigma_H^2\beta_0}\bg.
\end{equation}
Accordingly,
\begin{equation}
    \label{eq:app80}
  \sigma^2 = \bd^T\bbh^{-1}\bd = \sigma_H^2 -   \frac{\sigma_H^2\beta_0}{1+\sigma_H^2\beta_0}\sigma_H^2 =   \frac{\sigma_H^2}{1+\sigma_H^2\beta_0}.
\end{equation}
Substituting (\ref{eq:app79}) and (\ref{eq:app80}) to (\ref{eq:app76}), we finally obtain
\[
    \bv' = \bv + \frac{\delta}{1+\sigma_H^2\beta_1}(y-\hat y) \bg = \bv + \frac{\beta_1-\beta_0}{1+\sigma_H^2\beta_1}(y-\hat y)\bg.
\]

Equations (\ref{eq:update_zA}) and (\ref{eq:update_zC}) can be established by direct substitutions of (\ref{eq:update_v}) into their definitions (\ref{eq:def_bzA}) and (\ref{eq:def_bzC}) and thus the proof of Theorem \ref{thm:1} is complete.
\subsection{Computation of critical points}
For ease of notation we work with $\rho$, defined by
\begin{equation}
    \label{eq:bea2rho}
  \rho = \frac{\beta_1-\beta_0}{1+\sigma_H^2\beta_1}.
\end{equation}
It is easy to see that over the range $\beta_1>\beta_0$, $\rho$ is monotonically increasing in $(0,1/\sigma_H^2)$. Therefore, (\ref{eq:bea2rho}) can be inverted by
\begin{equation}
    \label{eq:rho2bea}
  \beta_1 = \frac{\rho+\beta_0}{1-\sigma_H^2\rho},
\end{equation}
where $\rho \in (0,1/\sigma_H^2)$ to ensure $\beta_1 > \beta_0$.

Suppose we have obtained $\rho^{(k)}, k=1,...,4$, $\beta_1^{(k)}$ can be calculated using (\ref{eq:rho2bea}) and the critical point $\beta_1$ is then
\[
    \beta_1 = \min_{k=1,...,4} \beta_1^{(k)}.
\]

We now calculate the critical value of $\rho$ for each condition one by one.
\subsubsection{Critical point for Condition 1}
Define the temporary vector
\[
\bt_\A= \left( {y - \hat y} \right)\left\{ {{{\bf{x}}_\A} -
\bl \bbbr_{\A\A}\bbs~~\bbbr_{\A\B}\br
\bg} \right\}.
\]
According to (\ref{eq:update_zA}),
\[
\la \bz_\A'= \la \bz_\A+ \rho\bt_\A.
\]
Condition 1 is met for any $\rho = \rho_i^{(1)}$ such that
\[
\rho_i^{(1)} = -\frac{\la z_{i}}{t_{i}}, i \in \A.
\]
Therefore, the critical value of $\rho$ that satisfies Condition 1 is
\[
\rho^{(1)} = \min\blc \rho_i^{(1)}\left|i \in \A, \rho_i^{(1)}  \in (0, 1/\sigma_H^2)\right.\brc.
\]
\subsubsection{Critical point for Condition 2}
By the definition (\ref{eq:def_v}), $\bv$ is a concatenation of $\alpha_m$ and $\bw_{\B_m}, m\in \PP$:
\begin{equation}
    \label{eq:strcutre_v}
\bv^T = \bl(\alpha_m)_{m\in\PP}, \bw_{\B_1}^T, ..., \bw_{\B_{|\PP|}}^T\br,
\end{equation}
where $(\alpha_m)_{m\in\PP}$ denotes the vector comprised of $\alpha_m, m\in \PP$.
Now we partition the  vector $\bg$ in the same manner as (\ref{eq:strcutre_v}) and denote $\tau_m$ and $\bu_m$ as the counter part of $\alpha_m$ and $\bw_{\B_m}$ in $\bg$, \ie
\[
\bg^T=\bl(\tau_m)_{m\in\PP}, \bu_1^T, ..., \bu_{|\PP|}^T\br.
\]
Eq. (\ref{eq:update_v}) is then equivalent to
\begin{equation}
    \label{eq:alphaprime}
  \alpha_m' = \alpha_m + \rho \tau_m,
\end{equation}
and
\[
  w_{\B_m,i}' =  w_{\B_m,i} +\rho u_{m,i},
\]
where $u_{m.i}$ is the $i$-th element of the vector $\bu_m$. Condition 2 indicates that
\[
\alpha_m' = \pm   w_{\B_m,i}',
\]
and is satisfied if $\rho= \rho_{m,i}^{(2+)}$ or $\rho = \rho_{m,i}^{(2-)}$, where
\[
\rho_{m,i}^{(2+)} = \frac{\alpha_m-w_{\B_m,i}}{u_{m,i}-\tau_m}, ~~ \rho_{m,i}^{(2-)} = -\frac{\alpha_m+w_{\B_m,i}}{u_{m,i}+\tau_m}.
\]
Therefore, the critical value of $\rho$ for Condition 2 is
\[
  \rho^{(2)} = \min \blc \rho_{m,i}^{(2\pm)} \left | m \in \PP, i = 1, ..., |\B_m|, \rho_{m,i}^{(2\pm)}  \in (0, 1/\sigma_H^2)
  \right.\brc.
\]
\subsubsection{Critical point for Condition 3}
According to (\ref{eq:alphaprime}), $\alpha_m' = 0$ yields $\rho = \rho^{(3)}_i$ determined by
\[
  \rho^{(3)}_m = -\frac{\alpha_m}{\tau_m}, m \in \PP,
\]
and the critical value for $\rho^{(3)}$ is
\[
  \rho^{(3)} = \min \blc \rho^{(3)}_m\left|, m \in \PP, \rho_m^{(3)}  \in (0, 1/\sigma_H^2) \right.
  \brc.
\]
\subsubsection{Critical point for Condition 4}
Define
\[
\bt_\C= \left( {y - \hat y} \right)\left\{ {{{\bf{x}}_\C} -
\bl \bbbr_{\C\A}\bbs~~\bbbr_{\C\B}\br
\bg} \right\}.
\]
Eq. (\ref{eq:update_zC}) is then
\[
  \la \bz_{\C_m}'= \la \bz_{\C_m}+ \rho\bt_{\C_m},
\]
and Condition 4 is equivalent to
\begin{equation}
    \label{eq:eq_condition4}
    \sum_{i\in\C_m} | \rho t_i+\la z_i| = \la.
\end{equation}
To solve (\ref{eq:eq_condition4}) we develop a fast method that requires complexity of $\OO(N\log N)$, where $N = |\C_m|$. The algorithm  is given in Appendix C. For each $m \in \Q$, let $\rho^{(4)}_m$ be the minimum positive solution to (\ref{eq:eq_condition4}). The critical value of $\rho$ for Condition 4 is then
\[
\rho^{(4)} = \min \blc \rho^{(4)}_m \left | m \in \Q, \rho_m^{(4)}  \in (0, 1/\sigma_H^2) \right.\brc.
\]
\subsection{Fast algorithm for critical condition 4}
Here we develop an algorithm to solve problem (\ref{eq:eq_condition4}). Consider solving the more general problem:
\begin{equation}
    \label{eq:weighted_abs}
\sum_{i = 1}^{N} a_{i}|x-x_{i}| = y,
\end{equation}
where $a_i$ and $x_i$ are constants and $a_i >0$. Please note that the notations here have no connections to those in previous sections. Define the following function
\[
h(x) = \sum_{i = 1}^{N} a_{i}|x-x_{i}|.
\]
The problem is then equivalent to finding $ h^{-1}(y) $, if it exists.

\begin{figure}
\centering
\includegraphics[width = 6cm]{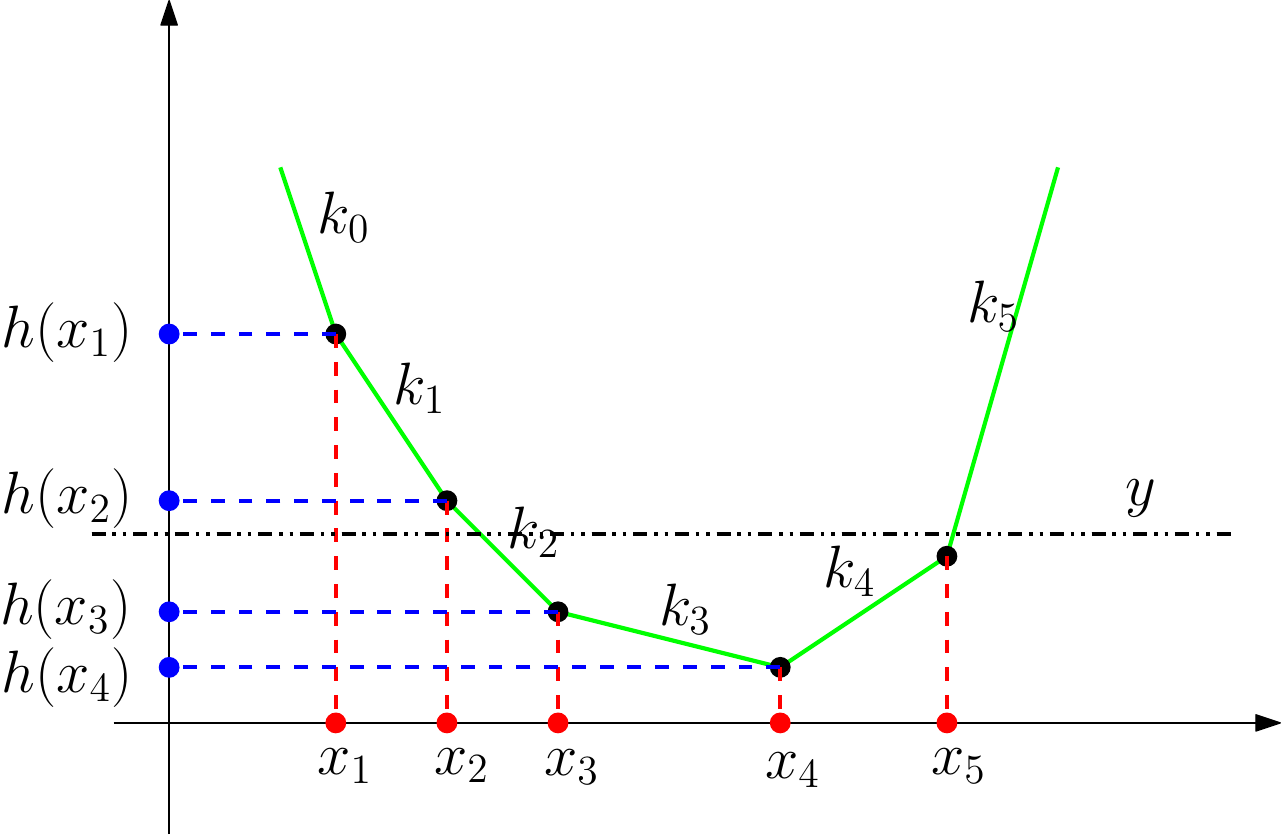}\\
  \caption{An illustration of the fast algorithm for critical condition 4.}\label{fig:app_1}
\end{figure}

An illustration of the function $h(x)$ is shown in Fig. \ref{fig:app_1}, where $k_i$ denotes the slope of the $i$th segment. It can be shown that $h(x)$ is piecewise linear and convex in $x$. Therefore, the equation (\ref{eq:weighted_abs}) generally has two solutions if they exist, denoted as $x_{\min}$ and $x_{\max}$. Based on piecewise linearity we propose a search algorithm to solve (\ref{eq:weighted_abs}). The pseudo code is shown in Algorithm \ref{alg:2} and its computation complexity is dominated by the sorting operation which requires $\OO(N\log N)$.

\begin{algorithm}
\label{alg:2}
\SetKwInOut{Input}{Input}\SetKwInOut{Output}{output}
\Input{$\{a_i, x_i\}_{i=1}^N$, $y$}
\Output{$x_{\min}, x_{\max}$}
\BlankLine

Sort $\{x_i\}_{i=1}^N$ in the ascending order: $ x_1 \le x_2 \le ... \le x_N $\; Re-order $\{a_i\}_{i=1}^N$ such that $a_i$ corresponds to $x_i$\;
Set $k_0 = - \sum_{i=1}^{N} a_i$\;
\For{$i=1,...,N$}
{
    $ k_i = k_{i-1} + 2a_{i}$\;
}
Calculate $h_1 = \sum_{i=2}^N a_i |x_1-x_i|$\;
\For{$i=2,...,N$}
{
    $h_i = h_{i-1} + k_{i-1}(x_i-x_{i-1})$
}
\eIf{ $\min_i k_i > y$ }
{
    No solution\;
    \textbf{Exit}\;
}
{
    \eIf{ $y > h_1$}
    {
        $x_{\min} = x_1 + (y-h_1)/k_0$\;
    }
    {
        Seek $j$ such that $y \in [h_j, h_{j-1}]$\;
        $x_{\min} = x_j + (y-h_j)/k_{j-1}$\;
    }
    \eIf{ $y > h_N$}
    {
        $x_{\max} = x_N + (y-h_N)/k_N$\;
    }
    {
        Seek $j$ such that $y \in [h_{j-1}, h_j]$\;
        $x_{\max} = x_{j -1}+ (y-h_{j-1})/k_{j-1}$\;
    }
}
  \caption{Solve $x$ from $\sum_{i = 1}^{N} a_{i}|x-x_{i}| = y$.}
\end{algorithm}

\bibliographystyle{IEEEbib}
\bibliography{refs}
\end{document}